\documentclass[5p,twocolumn]{elsarticle}
\usepackage{graphicx,latexsym}
\usepackage{dcolumn}
\usepackage{amssymb,amsmath,bm}
\usepackage{subfigure}
\usepackage{braket}
\usepackage{siunitx}
\usepackage{array}
\usepackage{float}
\usepackage[nopar]{lipsum}
\usepackage{caption}
\usepackage{multirow}
\usepackage{bigstrut}
\usepackage[super]{nth}

\usepackage{hyperref}
\hypersetup{
    pdfnewwindow=true,       
    colorlinks=true,         
    linkcolor=blue,          
    citecolor=blue,          
    filecolor=magenta,       
    urlcolor=black           
}

\usepackage[normalem]{ulem}

\def\sec#1{Sec.\ \ref{#1}}
\def\eq#1{Eq.\ (\ref{#1})}
\def\fig#1{Fig.\ \ref{#1}}

\journal{}

\begin{document}

\begin{frontmatter}


\title{Single photon controlled steady state electron transport through a resonance DQD-Cavity system in a strong coupling regime}

\author[a1,a2]{Halo Anwar Abdulkhalaq}
\ead{halo.abdulkhalaq@univsul.edu.iq}
\address[a1]{Division of Computational Nanoscience, Physics Department, College of Science, \\
	University of Sulaimani, Sulaimani 46001, Kurdistan Region, Iraq}
\address[a2]{Computer Engineering Department, College of Engineering, \\ Komar University of Science and Technology, Sulaimani 46001, Kurdistan Region, Iraq}

\author[a1,a2]{Nzar Rauf Abdullah}
\ead{nzar.r.abdullah@gmail.com}

\author[a4]{Vidar Gudmundsson}
\address[a4]{Science Institute, University of Iceland, Dunhaga 3, IS-107 Reykjavik, Iceland}
\ead{vidar@hi.is}


\begin{abstract}

We perform theoretical calculations to study steady-state electron transport in a double quantum dot, DQD, coupled to a quantized cavity photon field both in resonance and off-resonance regimes considering weak and strong coupling. In the resonant strong coupling regime, photon exchanges between the energy states of the DQD and the cavity are found reflecting multiple Rabi-resonances. The electron occupation of the states and the transport current can be smoothly increased by tuning the cavity-environment coupling strength.
Making the system off-resonant, but still in the strong coupling regime, the photon exchange is diminished and the electron occupation of the system and the transport current through it are prominent for high cavity-environment coupling strength.
In the weak coupling regime between the DQD and the cavity, the system has almost the same response in the resonant and the off-resonant regimes at high values of the cavity-environment coupling strength.

\end{abstract}

\begin{keyword}
Quantum dots \sep cavity quantum electrodynamics \sep quantum master equation
\end{keyword}

\end{frontmatter}

\section{Introduction}

The electron transport through quantum dot, QD, systems has been an interesting topic of research for many years. There are several techniques to control electron transport through QD systems
such as applied electric field \cite{Hauptmann2008}, magnetic field \cite{PhysRevB.82.195325, purcell1946resonance}, electrostatic gate voltage \cite{Hendrickx2018}, and application of external light sources \cite{Hsiao2020}.
The transport properties of QD systems can also be controlled by coupling the systems to single photon sources \cite{Hanschke:18}. These single photon sources have been considered for applications in the research fields of science and technology such as for efficient quantum memory \cite{Giannelli_2018} and optical quantum stimulators \cite{All-optical}.
In a QD system coupled to a photon cavity, spontaneous emission has been modified through the Purcell effect \cite{robin2005purcell, andre2006purcell}, local-field effects \cite{rahmani2001modification}, and the emission has been tuned by defects in a photonic crystal cavity \cite{spontaneous-englund2005controlling}. The Purcell effect enhancement is obtained by controlling the coupling of the cavity and the environment \cite{PhysRevLett.98.063601}. Moreover, the spontaneous emission rate can be enhanced
through geometrical design of the cavity \cite{cai2018photoluminescence, caligiuri2018planar,nano9050671}, or photon phase filters \cite{de2017solid}.

The different regimes of the cavity quantum electrodynamics (QED) couplings show variety of results, for instance a high impedance resonator enables a gate-defined DQD in strong coupling limit \cite{PhysRevX.7.011030}. The system is in strong coupling regime if the electron-photon coupling strength, $g_\gamma$, is greater than the cavity-environment coupling, $\kappa$, that is $g_\gamma>\kappa$, while in the weak coupling if $g_\gamma\leq\kappa$ \cite{ABDULKHALAQ2022115405}. These two coupling parameters $g_\gamma$ and $\kappa$ can be compared to see if the system is in the weak or the strong coupling regime \cite{snijders2018observation}. In the strong coupling regime, the results for plasmons and transition dipole moment coupling for the emitters are shown to lead to vacuum Rabi splitting in a single QD \cite{leng2018strong, dovzhenko2018light}. Moreover, a current peak with different types of photoluminescence is obtained due to the Rabi oscillations in a QD system strongly coupled to a quantized photon field \cite{lang2011observation, cirio2016ground}.

On the other hand, in the weak coupling regime some effects are observed such as an unconventional photon blockade in a single quantum
dot coupled to two polarized optical cavity modes \cite{snijders2018observation}. Furthermore, a photon number sensitivity is obtained in a two-level system using polarization postselection enabling high-quality photon sources \cite{snijders2018fiber, somaschi2016near}.

The study of the properties of electron transport assisted by a photon field in a QD system is one of the interesting aspects in the field of light-matter interaction \cite{gudmundsson2015coupled, childress2004mesoscopic}.
The electron transport through a QD system in the terahertz frequency range assisted multiphoton absorption up to the fourth order has been investigated \cite{shibata2012photon}. In addition, single-photons have been detected from the shot-noise in electron transport through a quantum point contact using a DQD \cite{gustavsson2007frequency}.

Motivated by the aforementioned studies, we investigate the electron transport through a DQD system.
The DQD system of our study is embedded in a short quantum wire coupled to a photon cavity and environment with the whole system placed in a uniform weak magnetic field. We study both the weak and the strong coupling regimes with the cavity-environment coupling, $\kappa$, tuned to different values compared to the electron-photon coupling, $g_\gamma$. Our investigation of the electron transport in the DQD system coupled to a photon cavity is performed in the steady state regime using a Markovian quantum master equation \cite{Efficient342017}.

The rest of the paper is organized as follow, in $\sec{Model_Theory}$ a brief description of the model and theoretical formalism is presented. Results obtained are discussed in $\sec{results}$, and in $\sec{conclusion}$ the concluding remarks are summarized.

\section{Method and Theory}\label{Model_Theory}

The system of our study is a double quantum dot (DQD) embedded in a short GaAs quantum wire with an effective mass $m^* = 0.067 m_{\rm e}$ and dielectric constant $\bar{\kappa} = 12.3$. The DQD-wire system is coupled to a quantized single mode photon cavity. The system is connected to two leads from left and right with chemical potential values $\mu_L=1.25$ meV and $\mu_L=1.15$ meV, respectively. The Leads potentials create a bias window in which some of the electron states can be located. In order to shift electron states into or out of this bias window a plunger gate voltage $V_g$ is used.
The DQD-wire system is a two-dimensional electron system with a hard-wall confined in the $x$-direction, the transport direction, at $x=\pm L_x/2$, and a parabolic confinement in the $y$-direction. The parallel quantum dots are symmetrical with the same diameter of $65$ nm, and the length of the quantum wire is $L_{x}=150$ nm.

The total system, the DQD system and the leads, is placed in a weak perpendicular external magnetic field $\textbf{B}=B\hat{\bm z}$ with magnitude of $0.1$ T. The effective confinement frequency of the magnetic field is given by $\Omega_w^2=w_c^2+\Omega_0^2$, where the $w_c=eB/m^*c$ is the cyclotron frequency. The effective magnetic length, $a_{\omega}=(\hbar/m^*\Omega_{\omega})^{1/2}$, is used to scale all lengths in the system. A schematic diagram of the model of our study is shown in \cite{ABDULKHALAQ2022414097}.

The total Hamiltonian, which describes the DQD system and the cavity many-body (MB), describes the electrons in the central system controlled by a plunger gate voltage and coupled to a single mode photon field. The total Hamiltonian includes the electron Hamiltonian $H_{\rm e}$ of the DQD system, the Hamiltonian of the photon field $H_{\rm \gamma}$, and  the Hamiltonian of the electron-photon coupling $H_{\rm e\text{-}\gamma}$.
The DQD system Hamiltonian is given by \cite{Rabi-resonant-abdullah2019}
\begin{equation}
	\begin{split}
		H_{\rm e} & =\sum_{ij}\bra{\Psi_{j}}\left[\frac{\pi_e^2}{2m^*}+eV_g+V_{\rm DQD}\right]\ket{\Psi_i}d_{j}^{\dagger}d_i \\
		& +H_Z+\frac{1}{2}\sum_{ijnm}V_{ijnm}d_{j}^{\dagger}d_{m}^{\dagger}d_id_n,
	\end{split}
	\label{H_e}
\end{equation}
where $|\Psi_i\rangle$ is a single-electron eigenstate (SES), $d_{i}^{\dagger}$ is the electron creation operator and $d_i$ is electron annihilation operator for the state $|\Psi_i\rangle$ of the central system. $\pi_e=\textbf{p}+\frac{e}{c}\textbf{A}_{ext}$, where $\textbf{p}$ is the momentum operator, $e$ the electron charge, and $\textbf{A}_{ext}$ is the vector potential of the external magnetic field \textbf{B} given by $\textbf{A}_{ext}=-B\hat{x}$. The Zeeman Hamiltonian, the second term of \eq{H_e}, is written as $H_Z=(g_s\mu_BB/2$)$\sigma_z$, where $g_s$ is the electron spin $g$-factor, $\mu_B$ is the Bohr magneton and $\sigma_z$ is spin-Pauli matrix. The third term of \eq{H_e} is the Coulomb interaction that contains the Coulomb integral given by
\begin{equation}
	V_{ijnm}=\bra{\Psi_{j}\Psi_{m}}\left[\frac{e^2}{\bar{\kappa}|r-r'|}\right]\ket{\Psi_i\Psi_n},
\end{equation}
where $\bar{\kappa}$ is dielectric constant, and $|r-r'|$ is  electron pair spatial separation.
The Hamiltonian of free photon field is written as $H_{\rm \gamma} = \hbar \omega_{\rm \gamma} a^{\dagger}a$, where $\hbar \omega_{\rm \gamma}$ is the photon energy, $a^{\dagger}$ and $a$ are the photon creation and annihilation operators, respectively. The Hamiltonian describing the electron-photon interaction is given by
\begin{equation}
	\begin{split}
		H_{\text e\text{-}\gamma} & =\frac{e}{m^*c} \sum_{ij}\bra{\Psi_{j}}{\bm \pi}_e\cdot\textbf{A}_{\gamma}\ket{\Psi_i}d_{j}^{\dagger}d_i \\
		& +\frac{e^2\textbf{A}_{\rm \gamma}^2}{2m^*c^2}
		\sum_{ij}\braket{\Psi_{j}|\Psi_i}d_{j}^{\dagger}d_i,
	\end{split}
\end{equation}
where $\textbf{A}_{\rm \gamma}$ is the quantized vector potential of the photon field given by $\textbf{A}_{\rm \gamma}=A(a+a^{\dagger}) \hat{\textbf{e}}$, in which $A$ is the amplitude of the photon field, $\hat{\textbf{e}}$ is the unit vector for x and y-polarization of the photon field given by $\hat{\textbf{e}}=(e_x,0)$ and $\hat{\textbf{e}}=(0,e_y)$ respectively. The electron-photon coupling strength is written in terms of the vector potential amplitude as $g_{\gamma}= A \Omega_w a_w/c$. We have assumed that the wavelength of the FIR (far-infrared) cavity field is much larger than the size of the DQD system. A numerically exact diagonalising technique is used for this calculation, using a tensor product of the Coulomb interacting MB basis states and eigenstates of the photon number operator \cite{2019b,Stepwise442012}.

In the steady state regime the Markovian master equation is used to calculating the electron transport through the DQD system coupled to the leads. The Markovian master equation is based on the projection of the dynamics of the total system onto the DQD system leading to a Nakajima and Zwanzig generalized master equation (GME) \cite{nakajima1958quantum, zwanzig1960ensemble}. The Markovian master equation is used to to calculate the reduced density operator $\rho_{\rm S}$, which describes the time evolution of the central system under the influence of the photon reservoir and the external leads \cite{correlations45b2018, vidar_2021}. The total density operator $\rho(t_0)$ as $\rho(t_0)=\rho_l(t_0)\rho_{\rm S}(t_0)$ is a tensor product of the uncorrelated reduced density operators of the central system $\rho_S(t_0)$ and external leads $\rho_{\rm l}(t_0)$, before the coupling of central system to
the external leads, where $t_0$ represents a time before the coupling. The reduced density operator of the central system, after the coupling at $t>t_0$, is written as \cite{Efficient342017}
\begin{equation}
	\rho_{\rm S}(t)=Tr_{L,R}[\rho],
\end{equation}
where $L$ and $R$ represent the left and the right leads of the electron reservoirs.
Now, the total current from left to right through the DQD system can be calculated from the reduced density operator as
\begin{equation}
	I_{L,R}=Tr[\rho_{\rm S}^{L,R}(t) \, Q]
\end{equation}
where $\rho_S^{L,R}(t)$ is reduced density operator, and $Q=-e\sum_{n}d_n^{\dagger}d_n$ is the charge operator of the DQD system  \cite{gudmundsson2017regimes, hohenester2010cavity}.

\section{Results}\label{results}

In this section, we present the results of the calculations of the electron transport through the DQD system. We study the transport properties in a DQD system by tuning the cavity-environment coupling strength, $\kappa$, into both the strong and the weak coupling regimes.
The DQD system of our study are two parallel symmetrical quantum dots embedded in a short quantum wire of length $L=150$~nm which is connected to two electron reservoirs (leads) from the left and the right. The two leads at a constant temperature $T=0.5$~K have chemical potentials $\mu_L=1.25$ meV and $\mu_R=1.15$ meV.
The system is coupled to a single mode photon cavity with one photon initially present in the cavity, n$_{\rm R}=1$, and the system is placed in a weak external magnetic field with strength $B=0.1$ T.

The gate voltage is fixed at $eV_g=0.6$ eV, while the photon energy is $\hbar\omega_{\rm \gamma}=0.6$ meV, which makes the system to be in resonance. For the values $\hbar\omega_{\rm \gamma}=0.45$ and 0.75 meV the system is off-resonant.
The resonance is between the ground state and the first-excited state.
The shown results will be for the $x$-polarization of the photon field since the $y$-polarization leads almost to the same characteristic as the $x$-polarization \cite{ABDULKHALAQ2022414097}.
\subsection{Strong coupling regime, $g_{\gamma}>\kappa$}\label{weak}

\begin{figure}[htb]
	\centering
	\includegraphics[width=0.5\textwidth]{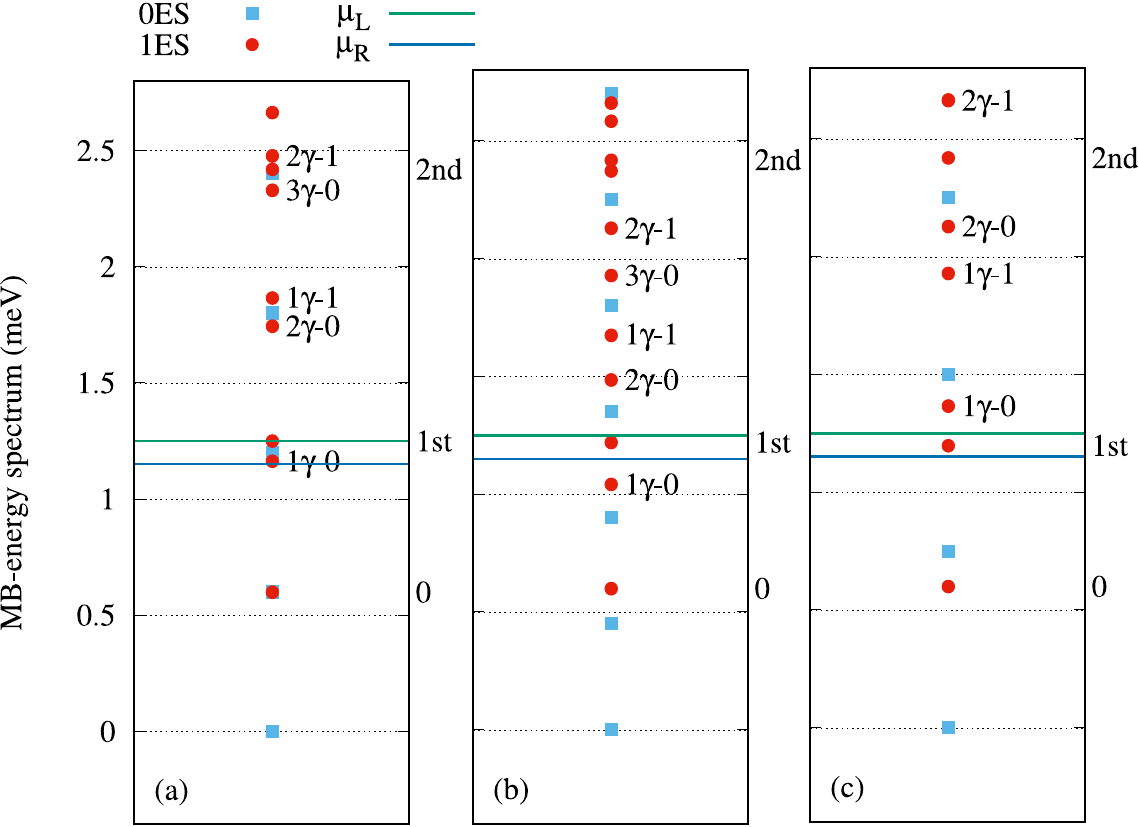}
	\caption{Many-Body (MB) energy spectrum for some selected states of a DQD system coupled to a photon cavity for three different values of the photon energy (a) $\hbar\omega_{\rm \gamma}=0.6$ meV (resonance), (b) $0.45$ meV (off-resonance-I) and (c) $0.75$ meV (off-resonance-II), where 0ES (blue squares) represent zero-electron states, and 1ES (red circles) are one-electron states. The green and purple lines are the chemical potential of left lead, $\mu_L=1.25$ meV, and right lead, $\mu_R=1.15$ meV, respectively. 0 labels the one-electron ground-state energy, while 1 and 2 represent the one-electron first- and second-excited states, respectively. 1$\gamma$-0, 2$\gamma$-0, and 3$\gamma$-0, respectively, refer to the first, the second, and the third photon replica of the ground state. The initial number of cavity photon is $n_R=1$, the electron-photon coupling strength is $g_{\gamma}=0.15$ meV  and the cavity-environment coupling strength is $\kappa=10^{-5}$ meV. The system is placed in a weak external magnetic field $B=0.1$ T, the plunger gate voltage is $V_g=0.6$ meV, the leads temperature is $T_{L,R}=0.5$ K. }
	\label{fig01}
\end{figure}

In this case the electron-photon coupling, $g_{\gamma}=0.15$ meV, is considered to be larger than or comparable to the selected range of the cavity-environment coupling strength $\kappa$.
In \fig{fig01}, the MB-energy spectrum of the DQD-cavity system is shown for the three values of photon energy (a) $\hbar\omega_{\rm \gamma}=0.6$ meV (resonance), (b) $0.45$ meV (off-resonance-I), and (c) $0.75$ meV (off-resonance-II). Here, the electron-photon coupling is larger than the cavity-environment coupling, hence the system is in the strong coupling regime.

Since the MB-energy spectrum is independent off $\kappa$, we only display the MBE
for one value of the cavity-environment coupling.
$0$, $1st$ and $2nd$ indicate the one-electron ground-, the first- and the second-excited states, respectively, and 1$\gamma$-0, 2$\gamma$-0 and 3$\gamma$-0 are, respectively, the first, the second and the third photon replica of the ground state. The green and blue horizontal lines represent the left and right leads chemical potential, respectively. In all three regimes (resonance, off-resonance-I and off-resonance-II) the first excited state is located in the bias window, while the ground and the second excited state are below and above the bias window, respectively.

\begin{figure}[htb]
	\centering
	\includegraphics[width=0.45\textwidth]{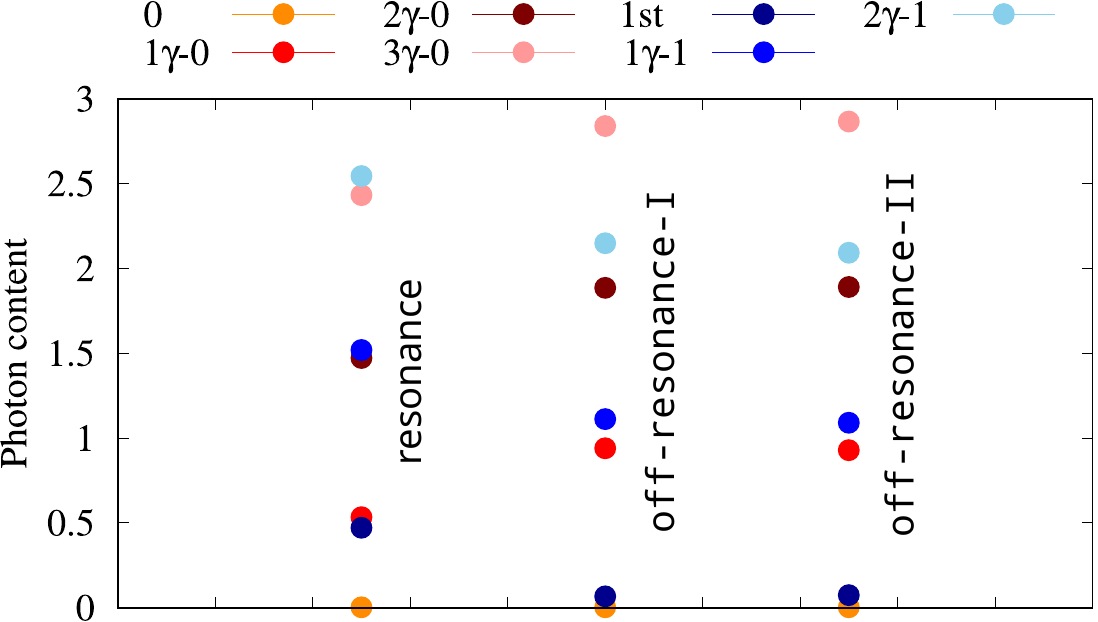}
	\caption{Photon content for some selected states of the DQD system coupled to a photon cavity for three different values of the photon energy (a) $\hbar\omega_{\rm \gamma}=0.6$ meV (resonance), (b) $0.45$ meV (off-resonance-I) and (c) $0.75$ meV (off-resonance-II). $0$ is the one-electron ground-state, while $1st$ is the first-excited state. 1$\gamma$-0, 2$\gamma$-0, and 3$\gamma$-0, respectively, refer to the first, the second, and the third photon replica of the ground state. The initial cavity photon number is $n_R=1$, the electron-photon coupling strength is $g_{\gamma}=0.15$ meV  and $\kappa=10^{-5}$ meV. The system is placed in a weak external magnetic field $B=0.1$ T, the plunger gate voltage is $V_g=0.6$ meV, the leads temperature is $T_{L,R}=0.5$ K. }
	\label{fig02}
\end{figure}
In the resonance regime, where the photon energy is almost equal to the energy spacing between the ground and first-excited states, the first photon replica of the ground state, 1$\gamma$-0, is inside the bias window as is the first-excited state, $1st$. To confirm the resonances, the photon content of the states is presented in \fig{fig02}. It can be clearly seen that the 1$\gamma$-0 shares its photon content with the $1st$ state, and both states thus have almost the same photon content, $0.5$. In the same way, there is photon sharing process between $2\gamma\text{-}0$ and $1\gamma\text{-}1$ with a photon content of $1.5$, and the $3\gamma\text{-}0$ and $2\gamma\text{-}1$ with a photon content of $2.5$. This indicates a multiple resonance process between the states at $\hbar\omega_{\rm \gamma}=0.6$ meV.

On the other hand, in the off-resonance-I regime the 1$\gamma$-0 is located below the bias window, while in off-resonance-II regime it is above the bias window (see \fig{fig01}(b,c)). In these two off-resonant cases the photon energy is not equal to the energy spacing between $0$ and $1st$.
Consequently, only a very small photon sharing process is seen between the states (see \fig{fig02}).
The photon content of the states does not depend on the value of cavity-environment coupling strength.

\begin{figure}[htb]
	\centering
	\includegraphics[width=0.4\textwidth]{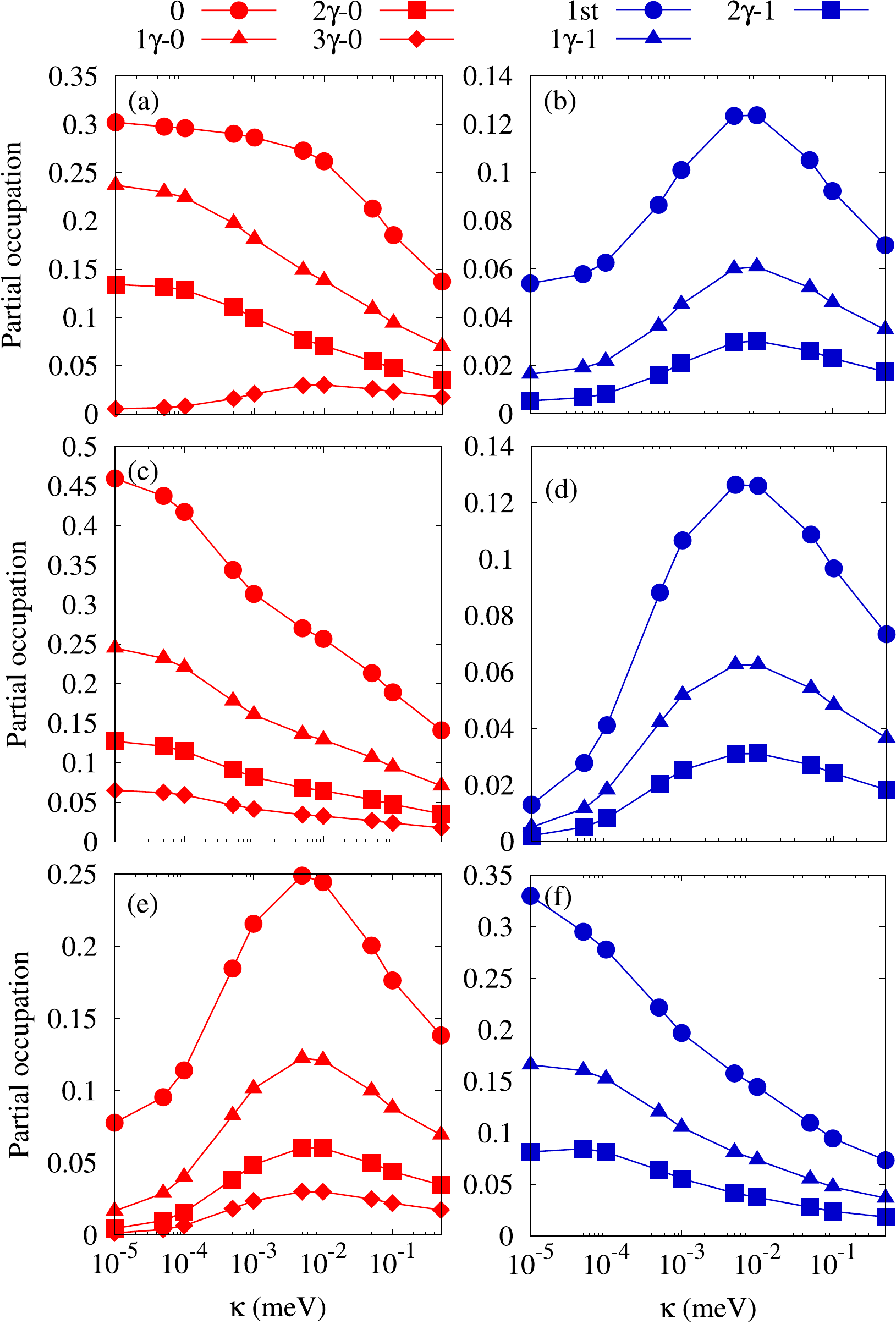}
	\caption{Partial occupation as a function of $\kappa$ for the ground and its photon replica states (red) at different photon energies (a) $\hbar\omega_{\rm \gamma}=0.6$, (c) $0.45$, (e) $0.75$~meV, and the first-excited state and its photon replica states (blue) (b) $\hbar\omega_{\rm \gamma}=0.6$, (d) $0.45$, (f) $0.75$~meV, in the strong coupling regime where $g_{\gamma}=0.15$ meV in case of $x$-polarized photon field. The initial photon number in the cavity is $n_R=1$, $B=0.1$ T, $V_g=0.6$ meV, and $T_{L,R}=0.5$ K.}
	\label{fig03}
\end{figure}

One of the most important parameters to understand the transport properties of the DQD is the partial or the total electron occupation of the system.
\fig{fig03} displays the partial occupation as a function $\kappa$ for the ground and its photon replica states (red) at different photon energies including (a) $\hbar\omega_{\rm \gamma}=0.6$, (c) $0.45$, (e) $0.75$~meV, and first-excited state and its photon replica states (blue) (b) $\hbar\omega_{\rm \gamma}=0.6$, (d) $0.45$, (f) $0.75$~meV.

In the resonant regime, $\hbar\omega_{\rm \gamma}=0.6$~meV (see \fig{fig03}(a,b)), the occupation of the ground state, $0$, is higher than that of the first-excited state, $1st$, at the low value of $\kappa=10^{-5}$, which is expected as the $0$ state is located below the Fermi energy of the leads. The occupation of the replicas of the ground state follow the $0$ state, and the replicas of $1st$ have almost the same qualitative characteristics as $1st$ state. Even though the $1st$ state and the $1\gamma\text{-}0$ are in resonance they have different electron occupation at $\kappa =10^{-5}$. This indicates that the photon lifetime in the vacuum of the cavity is long at this low value of $\kappa$ and the photon absorption process by the electrons is slow. Increasing the value of $\kappa$, the $1st$ is populated while the $1\gamma\text{-}0$ state is depopulated, and the occupation of both states reaches almost the same value at $\kappa = 10^{-2}$. This demonstrates a strong Rabi-resonance between the $1st$ and the $1\gamma\text{-}0$ states. Increasing further the value of $\kappa$ in which $g_{\gamma} \approx \kappa$, both the $1st$ and $0$ states are depopulated, which is due to increasing inelastic photon processes in the cavity weakening the Rabi-resonances.

In the off-resonance-I case, the $1\gamma\text{-}0$ is located below the bias window (see \fig{fig01}(b)), which is similar to the resonance case where the $1\gamma\text{-}0$ is below the
$1st$ state but staying in the bias window. The difference here is that there is a very small photon exchange between the states (see \fig{fig02}) in the case of off-resonance-I indicating a very weak Rabi-resonance between the states.
Consequently, we see that the occupation of the $0$ state and its photon replica states is much higher than that of the $1st$ state and it's photon replica states at $\kappa = 10^{-5}$ shown in \fig{fig03}(c,d).
By increasing the value of $\kappa$, the photon content of the states remains unchanged but we see that the electron occupation character of the states approaches the occupation seen in the resonance case. 

In the case of off-resonance-II regime, the partial occupation of the states is in many ways
totally opposite to the off-resonance-I case (see \fig{fig03}(e,f)), which is expected as the $1\gamma\text{-}0$ is now located above the bias window and even above the $1st$.

\begin{figure}[htb]
	\centering
	\includegraphics[width=0.4\textwidth]{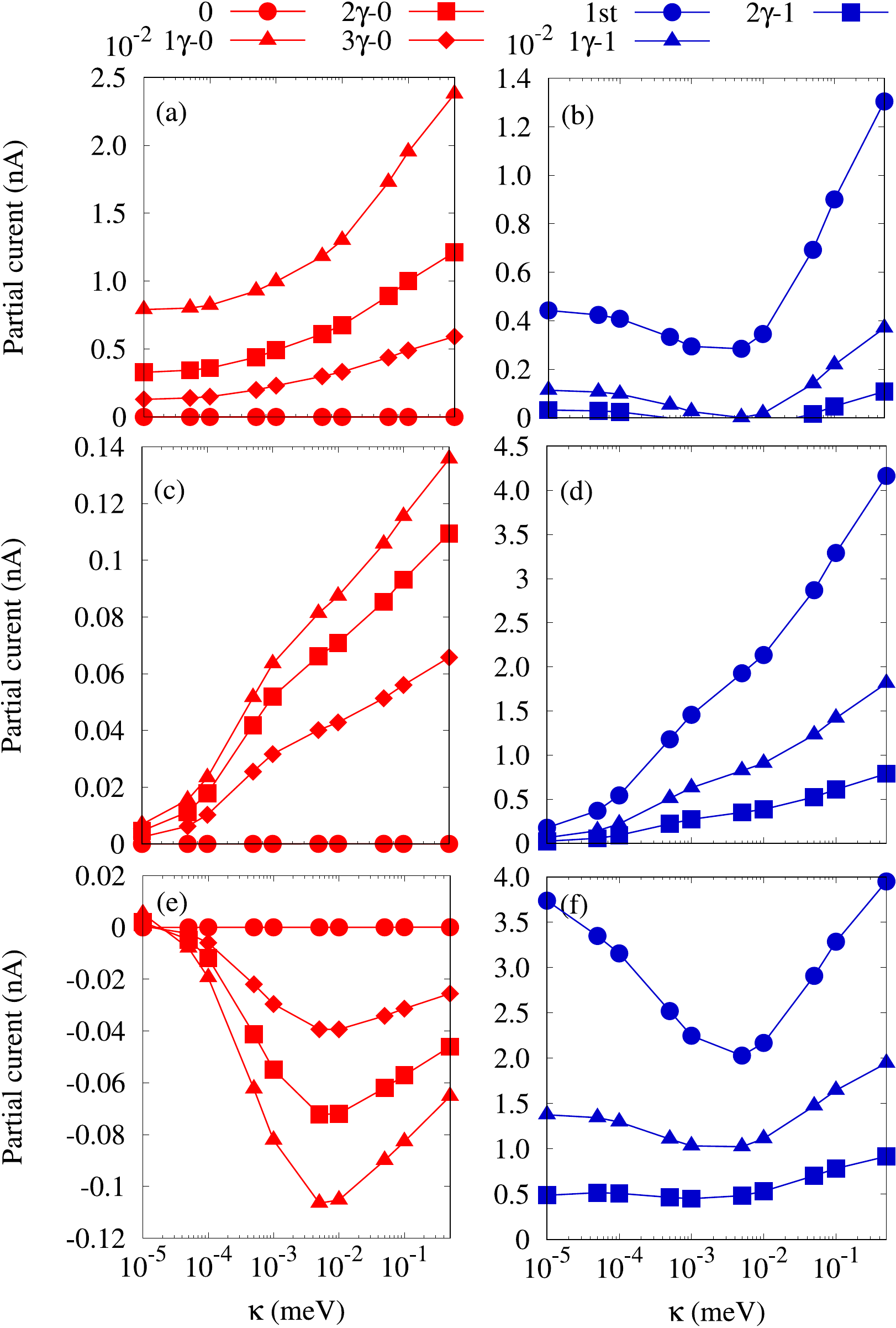}
	\caption{Partial current from the left lead to the DQD system as a function of the cavity-environment coupling strength $\kappa$ for the $0$ state and its photon replica states (red) at different photon energies of (a) $\hbar\omega_{\rm \gamma}=0.6$ meV, (c) $0.45$, (e) $0.75$~meV, and the $1st$ state and its photon replica states (blue) at (b) $\hbar\omega_{\rm \gamma}=0.6$, (d) $0.45$, (f) $0.75$~meV, in the strong coupling regime where the value of $g_{\gamma}=0.15$ meV in the case of a $x$-polarized photon field. The initial photon number in the cavity is $n_R=1$, $B=0.1$ T, $V_g=0.6$ meV and $T_{L,R}=0.5$ K.}
	\label{fig04}
\end{figure}

To further explore the transport properties of the DQD system, we now look at the partial current of the mentioned selected states. The partial current as a function of $\kappa$ in the strong coupling regime, with $g_{\gamma}=0.15$ meV and $g_{\gamma} > \kappa$, is shown in \fig{fig04}. The partial current is displayed for the $0$ state and its photon replica states (red) and the $1st$ state and its photon replica states (blue) at different photon energies, which are (a,b) $\hbar\omega_{\rm \gamma}=0.6$, (c,d) $0.45$, and (e,f) $0.75$~meV.

In the resonant regime (see \fig{fig04}(a,b)), we see that the $0$ state is charged but the current via the $0$ state is almost zero because it is located well below the bias window indicating that the $0$ state does not contribute to the current transport to the DQD system. But the photon replica states of $0$ actively participate in the current due to the multiple resonances between the replicas of the $0$ state and the replicas of $1st$ state as follows: between $1\gamma\text{-}0$ and $1st$, between $2\gamma\text{-}0$ and $1\gamma\text{-}1$, and between $3\gamma\text{-}0$ and $2\gamma\text{-}1$. The most active states in the current transport are $1\gamma\text{-}0$ and $1st$, which is caused by the strong resonances among all the states and these two states are located in the bias window. A noticeable current reduction at $\kappa \approx 10^{-3}$ to $10^{-2}$ is seen for the $1st$ state and its photon replica states, while a current enhancement for the $0$ and its photon replica states in the same interval is found confirming the strong Rabi-resonance between these two states.

In the off-resonance-I case (see \fig{fig04}(c,d)), similar to the resonant regime the current via the $0$ state is zero, and the current via the replicas of the $0$ state is reduced compared to the resonant regime. This is because of the absence of, or very weak Rabi-resonance effect in the system. In this case, the $1st$ state is dominant in the current transport, which is due the location of the $1st$ in the bias window. It is clear that the replicas of the
$1st$ state directly follow the $1st$, and they thus have the same current characteristics as $1st$ state but with less weight.

Finally, in the off-resonance-II case (see \fig{fig04}(e,f)), the $0$ state has almost zero current and its photon replicas have reversed current transport directions, which means that the current goes from the DQD to the left lead as it has the negative value. The $1st$ and its photon replicas have positive value of current indicating the current going from left lead to the DQD. In the current transport here is a weak correlation between the $1st$ in the bias window and the $1\gamma\text{-}0$ state above the bias window. The current goes to the $1st$, and at the same time the $1\gamma\text{-}0$ try empty the DQD, which is because the $1\gamma\text{-}0$ is located just above the chemical potential of the left lead and there is an opportunity to transfer an electron from the DQD to the left lead. As a result, a negative current is seen for the $1\gamma\text{-}0$ state and other replicas of the $0$ state.

\begin{figure}[htb]
	\centering
	\includegraphics[width=0.5\textwidth]{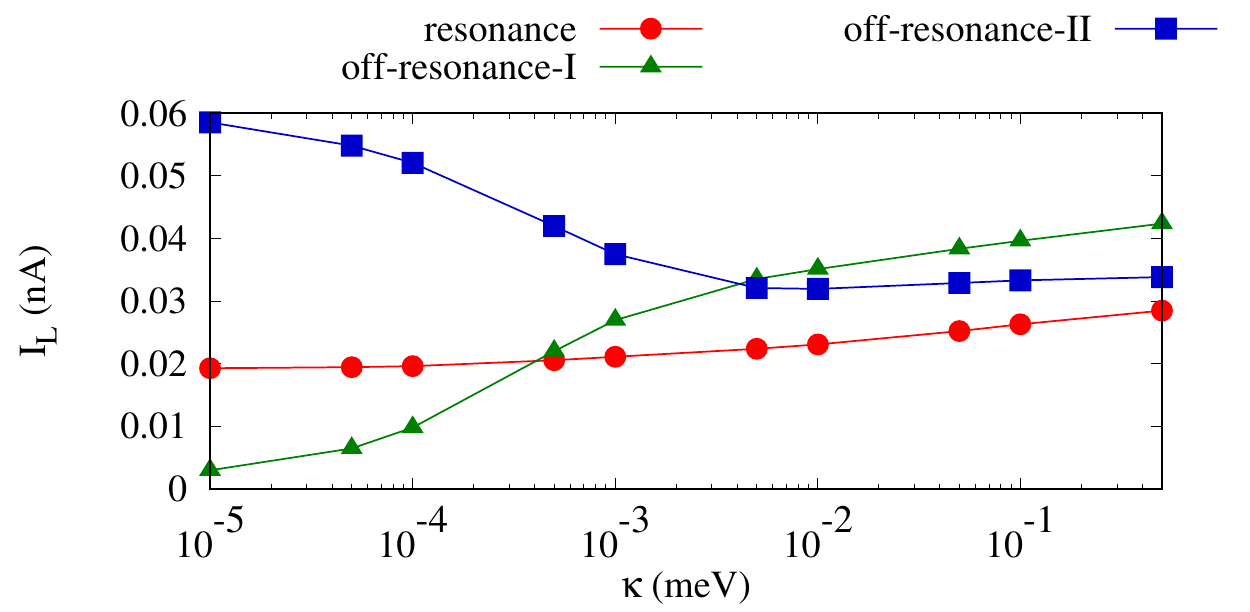}
	\caption{Total current from left lead to the DQD system as a function of $\kappa$ for the resonant regime at $\hbar\omega_{\rm \gamma}=0.6$~meV (red circle), the off-resonance-I case at $\hbar\omega_{\rm \gamma}=0.45$~meV (green triangle), and the off-resonance-II case at $\hbar\omega_{\rm \gamma}=0.75$~meV (blue squares). The system is in the strong coupling regime, where $g_{\gamma}=0.15$ meV for $x$-polarized photon field. The initial photon number in the cavity is $n_R=1$, $B=0.1$ T, $V_g=0.6$ meV and $T_{L,R}=0.5$ K.}
	\label{fig05}
\end{figure}

We now present the total left current in the DQD system, where the total current here is the total current going from the left lead to the DQD and vise versa. The current going from the DQD to the right lead has similar characteristics of the left lead because the system is in steady-state regime. We thus do not display the total current on the right side of the DQD system.
\fig{fig05} shows the total current from left lead to the DQD system as a function of $\kappa$ for resonant regime at photon energy $\hbar\omega_{\rm \gamma}=0.6$~meV (red circle), the off-resonance-I case at $\hbar\omega_{\rm \gamma}=0.45$~meV (green triangle), and the off-resonance-II case for $\hbar\omega_{\rm \gamma}=0.75$~meV (blue squares). The system is in the strong coupling regime, where $g_{\gamma}=0.15$~meV. The behavior of the total current in the resonance case follows the partial currents of the most contributing states, which are the $0$ state and its photon replicas, and the $1st$ state and its photon replica states as is seen in \fig{fig04}(a,b). The total current smoothly increases with $\kappa$, and the total current shows a slow increase compared to the current of two off-resonance regimes. The total current for the off-resonance-I case is almost zero for very low value of $\kappa = 10^{-5}$, which is because both the $0$ and the $1st$ states and their replicas have almost zero current, and then the total current is increases as $\kappa$ is increased.

In contrast, the total current decreases with $\kappa$ for the off-resonance-II case.
At the low value of $\kappa = 10^{-5}$, the total current has the highest value which is due to the high partial current of the $1st$ and its photon replica states. The total current decreased with $\kappa$ up to $\kappa =10^{-2}$, which is due to the negative value of current of the photon replica state of $0$.

In general, the total current has different properties for both the resonant and the off-resonance regimes for low values of $\kappa$, even it has different character for the different off-resonance regimes.
But for high values of $\kappa$, the total current has a plateau characteristics for all the considered regimes and the total current does not change much with $\kappa$. In addition, in the off-resonance regimes the total current is enhanced at the high value of $\kappa$ compared to total current in the resonant regime.  We can also confirm that the characteristics of the occupation, the partial current, and the total current are qualitatively unchanged if the strength of electron-photon coupling is further increased, but we have to assume that the simple structure of the dissipation terms may not be adequate any more in that regime.
\subsection{Weak coupling regime,  $g_{\gamma}<\kappa$}\label{strong}

\begin{figure}[htb]
	\centering
	\includegraphics[width=0.5\textwidth]{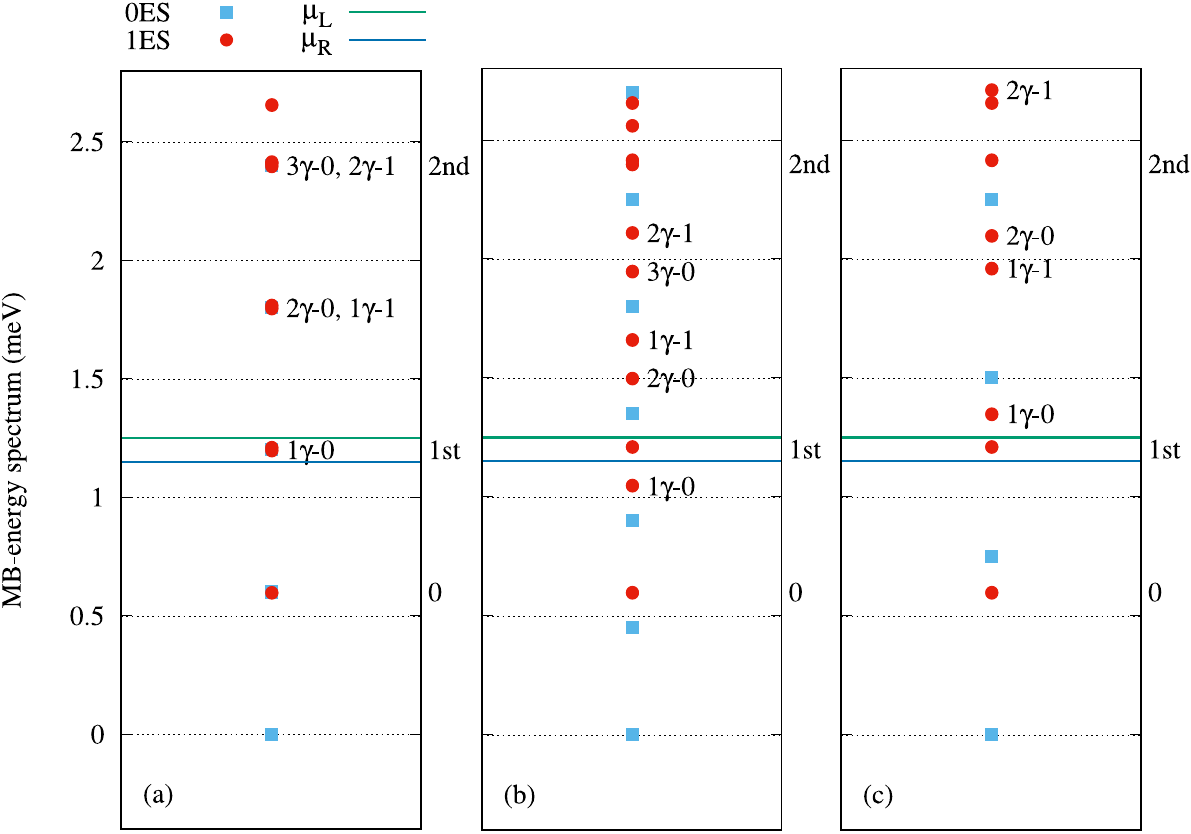}
	\caption{Many-Body (MB) energy spectrum of the DQD system coupled to a photon cavity for three different values of the photon energy (a) $\hbar\omega_{\rm \gamma}=0.6$ (resonance), (b) $0.45$ (off-resonance-I) and (c) $0.75$ meV (off-resonance-II) in the weak coupling regime, $g_{\gamma} = 10^{-4}$~meV. The 0ES (blue squares) represent zero-electron states, and 1ES (red circles) are one-electron states. The green and purple lines are the chemical potential of left lead, $\mu_L=1.25$ meV, and right lead, $\mu_R=1.15$ meV, respectively. 0 is one-electron ground-state energy, while 1 and 2 are one-electron first- and second-excited state respectively. the 1$\gamma$-0, 2$\gamma$-0, and 3$\gamma$-0, respectively, refer to the first, the second, and the third photon replica of ground state. The initial cavity photon number is $n_R=1$, and the cavity-environment coupling strength is $\kappa=10^{-5}$ meV. The system is placed in a weak external magnetic field $B=0.1$ T, the plunger gate voltage is $V_g=0.6$ meV, the leads temperature is $T_{L,R}=0.5$ K. }
	\label{fig06}
\end{figure}

In this section, we set the electron-photon coupling to be $g_{\gamma}=10^{-4}$ meV,
which is comparable to or smaller than the selected range of $\kappa$.
In \fig{fig06}, the MB-energy spectrum of the DQD-cavity system is shown for three values of the photon energy (a) $\hbar\omega_{\rm \gamma}=0.6$ (resonance), (b) $0.45$ (off-resonance-I),
and (c) $0.75$~meV (off-resonance-II). The $0$, $1st$ and $2nd$ indicate
the one-electron ground, the first- and the second-excited state, respectively,
and 1$\gamma$-0, 2$\gamma$-0 and 3$\gamma$-0 are, respectively, the first, the second and the third photon replica states of the $0$ state.
The green and blue horizontal lines represent the chemical potential of the left and the right leads, respectively.
In the resonant regime, compared to the strong coupling regime as in \fig{fig01}(a), we can see that
the $1st$ and 1$\gamma$-0 states coincide indicating a weaker Rabi-splitting, but still both states are staying inside the bias window. Similar coincidence of the states is seen for the other photon replica states such as 2$\gamma$-0 with 1$\gamma$-1, and 2$\gamma$-1 with 3$\gamma$-0.
This demonstrates that the Rabi-resonances do not play an important role here due to the small value of $g_{\gamma}$.

\begin{figure}[htb]
	\centering
	\includegraphics[width=0.45\textwidth]{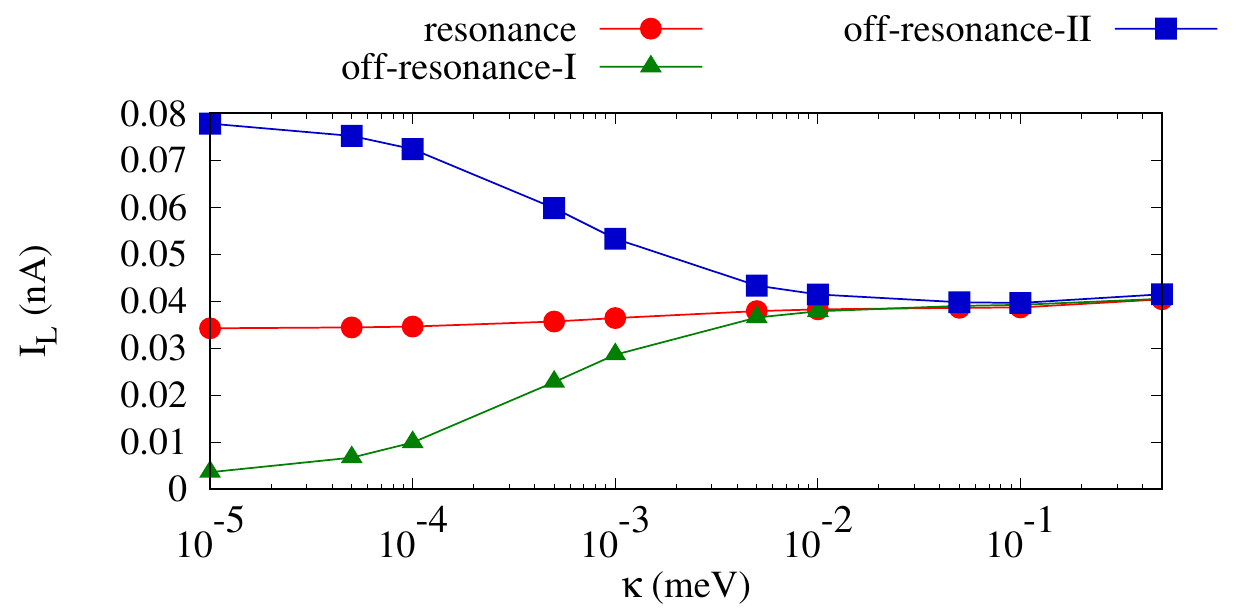}
	\caption{ Total current from left lead to the DQD system as a function of the cavity-environment coupling strength $\kappa$ for photon energies at resonance (red circle) $\hbar\omega_{\rm \gamma}=0.6$ meV, the off-resonance-I case (green triangle) $\hbar\omega_{\rm \gamma}=0.45$ and the off-resonance-II case (blue squares) $\hbar\omega_{\rm \gamma}=0.75$. The system is in the weak coupling regime, where the value of the electron-photon coupling $g_{\gamma}=10^{-4}$ meV for $x$-polarized photon field. The initial photon number in the cavity is $n_R=1$, $B=0.1$ T, $V_g=0.6$ meV and $T_{L,R}=0.5$ K. }
	\label{fig07}
\end{figure}

The weak coupling regime gives rise to a different current characteristics of the system.
In \fig{fig07}, the total current from left lead to the DQD system as a function of $\kappa$ is shown for the resonant regime at $\hbar\omega_{\rm \gamma}=0.6$ (red circle), the off-resonance-I case at $0.45$ (green triangle), and the off-resonance-II case at $0.75$~meV (blue squares). The system is in the weak coupling regime with $g_{\gamma}=10^{-4}$ meV.

In general, compared to total current in the strong coupling regime (\fig{fig05}) the value of the total current in the weak coupling regime is slighly increased, particularly that can clearly be seen in the resonant case (red circles). Similar to that of strong coupling, in both off-resonance regimes, the characteristics of the total current can be divided into two regions: $\kappa \geq g_{\gamma}$ and $\kappa < g_{\gamma}$. In the $\kappa \geq g_{\gamma}$ region, similar qualitative behavior to that of the strong coupling of total current can be seen only with smaller magnitude of the current. But in the $\kappa < g_{\gamma}$ region, all resonance and off-resonance regimes give almost the same qualitative and quantitative total current, which is due to the weak electron-photon coupling leading to the weak Rabi-resonance effects.

\section{Conclusion}\label{conclusion}

We have used a generalized master equation to investigate electron transport through a double quantum dot system resonantly or off-resonantly coupled to a single mode of a photon cavity in both the weak and the strong coupling regimes. The photon cavity is utilized to control or enhance the electric transport current through the DQD system. We find multiple Rabi-resonances in the system, which tend to increase or modify the electric current in the system.
The multiple-Rabi resonances are very weak in the off-resonant regimes leading to faster population of the DQD and the electric transport current is thus enhanced.
In addition, the effects of the electron-photon and the cavity-environment coupling strengths on the occupation and the electric current in the DQD system are discussed. Our results may be useful for optoelectronic devices in the nanoscale.

\section{Acknowledgment}
This work was financially supported by the University of Sulaimani and
the Research center of Komar University of Science and Technology.
The computations were performed on the IHPC facilities provided by the University of Iceland.



\begin{thebibliography}{10}
	\expandafter\ifx\csname url\endcsname\relax
	\def\url#1{\texttt{#1}}\fi
	\expandafter\ifx\csname urlprefix\endcsname\relax\def\urlprefix{URL }\fi
	\expandafter\ifx\csname href\endcsname\relax
	\def\href#1#2{#2} \def\path#1{#1}\fi
	
	\bibitem{Hauptmann2008}
	J.~R. Hauptmann, J.~Paaske, P.~E. Lindelof,
	\href{https://doi.org/10.1038/nphys931}{Electric-field-controlled spin
		reversal in a quantum dot with ferromagnetic contacts}, Nature Physics 4~(5)
	(2008) 373--376.
	\newblock \href {https://doi.org/10.1038/nphys931}
	{\path{doi:10.1038/nphys931}}.
	\newline\urlprefix\url{https://doi.org/10.1038/nphys931}
	
	\bibitem{PhysRevB.82.195325}
	N.~R. Abdullah, C.-S. Tang, V.~Gudmundsson,
	\href{https://link.aps.org/doi/10.1103/PhysRevB.82.195325}{Time-dependent
		magnetotransport in an interacting double quantum wire with window coupling},
	Phys. Rev. B 82 (2010) 195325.
	\newblock \href {https://doi.org/10.1103/PhysRevB.82.195325}
	{\path{doi:10.1103/PhysRevB.82.195325}}.
	\newline\urlprefix\url{https://link.aps.org/doi/10.1103/PhysRevB.82.195325}
	
	\bibitem{purcell1946resonance}
	E.~M. Purcell, H.~C. Torrey, R.~V. Pound, Resonance absorption by nuclear
	magnetic moments in a solid, Physical Review 69~(1-2) (1946) 37.
	
	\bibitem{Hendrickx2018}
	N.~W. Hendrickx, D.~P. Franke, A.~Sammak, M.~Kouwenhoven, D.~Sabbagh, L.~Yeoh,
	R.~Li, M.~L.~V. Tagliaferri, M.~Virgilio, G.~Capellini, G.~Scappucci,
	M.~Veldhorst,
	\href{https://doi.org/10.1038/s41467-018-05299-x}{Gate-controlled quantum
		dots and superconductivity in planar germanium}, Nature Communications 9~(1)
	(2018) 2835.
	\newblock \href {https://doi.org/10.1038/s41467-018-05299-x}
	{\path{doi:10.1038/s41467-018-05299-x}}.
	\newline\urlprefix\url{https://doi.org/10.1038/s41467-018-05299-x}
	
	\bibitem{Hsiao2020}
	T.-K. Hsiao, A.~Rubino, Y.~Chung, S.-K. Son, H.~Hou, J.~Pedr{\'o}s, A.~Nasir,
	G.~{\'E}thier-Majcher, M.~J. Stanley, R.~T. Phillips, T.~A. Mitchell, J.~P.
	Griffiths, I.~Farrer, D.~A. Ritchie, C.~J.~B. Ford,
	\href{https://doi.org/10.1038/s41467-020-14560-1}{Single-photon emission from
		single-electron transport in a saw-driven lateral light-emitting diode},
	Nature Communications 11~(1) (2020) 917.
	\newblock \href {https://doi.org/10.1038/s41467-020-14560-1}
	{\path{doi:10.1038/s41467-020-14560-1}}.
	\newline\urlprefix\url{https://doi.org/10.1038/s41467-020-14560-1}
	
	\bibitem{Hanschke:18}
	L.~Hanschke, K.~A. Fischer, S.~Appel, D.~Lukin, J.~J. Finley, J.~Vuckovic,
	K.~M\"{u}ller,
	\href{http://opg.optica.org/abstract.cfm?URI=CLEO_QELS-2018-FM1H.4}{Quantum
		dot single photon sources with ultra-low multi-photon error rate}, in:
	Conference on Lasers and Electro-Optics, Optica Publishing Group, 2018, p.
	FM1H.4.
	\newblock \href {https://doi.org/10.1364/CLEO_QELS.2018.FM1H.4}
	{\path{doi:10.1364/CLEO_QELS.2018.FM1H.4}}.
	\newline\urlprefix\url{http://opg.optica.org/abstract.cfm?URI=CLEO_QELS-2018-FM1H.4}
	
	\bibitem{Giannelli_2018}
	L.~Giannelli, T.~Schmit, T.~Calarco, C.~P. Koch, S.~Ritter, G.~Morigi,
	\href{https://doi.org/10.1088/1367-2630/aae725}{Optimal storage of a single
		photon by a single intra-cavity atom}, New Journal of Physics 20~(10) (2018)
	105009.
	\newblock \href {https://doi.org/10.1088/1367-2630/aae725}
	{\path{doi:10.1088/1367-2630/aae725}}.
	\newline\urlprefix\url{https://doi.org/10.1088/1367-2630/aae725}
	
	\bibitem{All-optical}
	S.~Cialdi, M.~A.~C. Rossi, C.~Benedetti, B.~Vacchini, D.~Tamascelli,
	S.~Olivares, M.~G.~A. Paris,
	\href{https://doi.org/10.1063/1.4977023}{All-optical quantum simulator of
		qubit noisy channels}, Applied Physics Letters 110~(8) (2017) 081107.
	\newblock \href {http://arxiv.org/abs/https://doi.org/10.1063/1.4977023}
	{\path{arXiv:https://doi.org/10.1063/1.4977023}}, \href
	{https://doi.org/10.1063/1.4977023} {\path{doi:10.1063/1.4977023}}.
	\newline\urlprefix\url{https://doi.org/10.1063/1.4977023}
	
	\bibitem{robin2005purcell}
	I.~Robin, R.~Andr{\'e}, A.~Balocchi, S.~Carayon, J.~G{\'e}rard, K.~Kheng, L.~S.
	Dang, H.~Mariette, S.~Moehl, F.~Tinjod, Purcell effect on cdse/znse quantum
	dots em bedded in pillar microcavities, physica status solidi (c) 2~(11)
	(2005) 3829--3832.
	
	\bibitem{andre2006purcell}
	R.~Andr{\'e}, I.~Robin, A.~Balocchi, S.~Carayon, S.~Moehl, J.~G{\'e}rard,
	Purcell effect on CdSe/ZnSe quantum dots in pillar microcavities, Physica
	Status Solidi (b) 243~(4) (2006) 827--830.
	
	\bibitem{rahmani2001modification}
	A.~Rahmani, G.~W. Bryant, Modification of spontaneous emission of quantum dots:
	Purcell effect in semiconductor microcavities, Physica Status Solidi (b)
	224~(3) (2001) 807--810.
	
	\bibitem{spontaneous-englund2005controlling}
	D.~Englund, D.~Fattal, E.~Waks, G.~Solomon, B.~Zhang, T.~Nakaoka, Y.~Arakawa,
	Y.~Yamamoto, J.~Vu{\v{c}}kovi{\'c}, Controlling the spontaneous emission rate
	of single quantum dots in a two-dimensional photonic crystal, Physical Review
	Letters 95~(1) (2005) 013904.
	
	\bibitem{PhysRevLett.98.063601}
	T.~Wilk, S.~C. Webster, H.~P. Specht, G.~Rempe, A.~Kuhn,
	\href{https://link.aps.org/doi/10.1103/PhysRevLett.98.063601}{Polarization-controlled
		single photons}, Phys. Rev. Lett. 98 (2007) 063601.
	\newblock \href {https://doi.org/10.1103/PhysRevLett.98.063601}
	{\path{doi:10.1103/PhysRevLett.98.063601}}.
	\newline\urlprefix\url{https://link.aps.org/doi/10.1103/PhysRevLett.98.063601}
	
	\bibitem{cai2018photoluminescence}
	Y.-Y. Cai, J.~G. Liu, L.~J. Tauzin, D.~Huang, E.~Sung, H.~Zhang, A.~Joplin,
	W.-S. Chang, P.~Nordlander, S.~Link, Photoluminescence of gold nanorods:
	Purcell effect enhanced emission from hot carriers, ACS NANO 12~(2) (2018)
	976--985.
	
	\bibitem{caligiuri2018planar}
	V.~Caligiuri, M.~Palei, M.~Imran, L.~Manna, R.~Krahne, Planar
	double-epsilon-near-zero cavities for spontaneous emission and Purcell effect
	enhancement, ACS photonics 5~(6) (2018) 2287--2294.
	
	\bibitem{nano9050671}
	W.~Wei, X.~Yan, J.~Liu, B.~Shen, W.~Luo, X.~Ma, X.~Zhang,
	\href{https://www.mdpi.com/2079-4991/9/5/671}{Enhancement of single-photon
		emission rate from InGaAs/GaAs quantum-dot/nanowire heterostructure by
		wire-groove nanocavity}, Nanomaterials 9~(5) (2019).
	\newblock \href {https://doi.org/10.3390/nano9050671}
	{\path{doi:10.3390/nano9050671}}.
	\newline\urlprefix\url{https://www.mdpi.com/2079-4991/9/5/671}
	
	\bibitem{de2017solid}
	L.~De~Santis, C.~Ant{\'o}n, B.~Reznychenko, N.~Somaschi, G.~Coppola,
	J.~Senellart, C.~G{\'o}mez, A.~Lema{\^\i}tre, I.~Sagnes, A.~G. White, et~al.,
	A solid-state single-photon filter, Nature Nanotechnology 12~(7) (2017)
	663--667.
	
	\bibitem{PhysRevX.7.011030}
	A.~Stockklauser, P.~Scarlino, J.~V. Koski, S.~Gasparinetti, C.~K. Andersen,
	C.~Reichl, W.~Wegscheider, T.~Ihn, K.~Ensslin, A.~Wallraff,
	\href{https://link.aps.org/doi/10.1103/PhysRevX.7.011030}{Strong coupling
		cavity QED with gate-defined double quantum dots enabled by a high impedance
		resonator}, Phys. Rev. X 7 (2017) 011030.
	\newblock \href {https://doi.org/10.1103/PhysRevX.7.011030}
	{\path{doi:10.1103/PhysRevX.7.011030}}.
	\newline\urlprefix\url{https://link.aps.org/doi/10.1103/PhysRevX.7.011030}
	
	\bibitem{ABDULKHALAQ2022115405}
	H.~A. Abdulkhalaq, N.~R. Abdullah, V.~Gudmundsson,
	\href{https://www.sciencedirect.com/science/article/pii/S1386947722002375}{Photon
		and magnetic field controlled electron transport of a multiply-resonant
		photon-cavity double quantum dot system}, Physica E: Low-dimensional Systems
	and Nanostructures 144 (2022) 115405.
	\newblock \href {https://doi.org/https://doi.org/10.1016/j.physe.2022.115405}
	{\path{doi:https://doi.org/10.1016/j.physe.2022.115405}}.
	\newline\urlprefix\url{https://www.sciencedirect.com/science/article/pii/S1386947722002375}
	
	\bibitem{snijders2018observation}
	H.~Snijders, J.~Frey, J.~Norman, H.~Flayac, V.~Savona, A.~Gossard, J.~Bowers,
	M.~Van~Exter, D.~Bouwmeester, W.~L{\"o}ffler, Observation of the
	unconventional photon blockade, Physical Review Letters 121~(4) (2018)
	043601.
	
	\bibitem{leng2018strong}
	H.~Leng, B.~Szychowski, M.-C. Daniel, M.~Pelton, Strong coupling and induced
	transparency at room temperature with single quantum dots and gap plasmons,
	Nature communications 9~(1) (2018) 1--7.
	
	\bibitem{dovzhenko2018light}
	D.~Dovzhenko, S.~Ryabchuk, Y.~P. Rakovich, I.~Nabiev, Light--matter interaction
	in the strong coupling regime: configurations, conditions, and applications,
	Nanoscale 10~(8) (2018) 3589--3605.
	
	\bibitem{lang2011observation}
	C.~Lang, D.~Bozyigit, C.~Eichler, L.~Steffen, J.~Fink, A.~Abdumalikov~Jr,
	M.~Baur, S.~Filipp, M.~P. Da~Silva, A.~Blais, et~al., Observation of resonant
	photon blockade at microwave frequencies using correlation function
	measurements, Physical Review Letters 106~(24) (2011) 243601.
	
	\bibitem{cirio2016ground}
	M.~Cirio, S.~De~Liberato, N.~Lambert, F.~Nori, Ground state
	electroluminescence, Physical Review Letters 116~(11) (2016) 113601.
	
	\bibitem{snijders2018fiber}
	H.~Snijders, J.~Frey, J.~Norman, V.~Post, A.~Gossard, J.~Bowers, M.~Van~Exter,
	W.~L{\"o}ffler, D.~Bouwmeester, Fiber-coupled cavity-QED source of identical
	single photons, Physical Review Applied 9~(3) (2018) 031002.
	
	\bibitem{somaschi2016near}
	N.~Somaschi, V.~Giesz, L.~De~Santis, J.~Loredo, M.~P. Almeida, G.~Hornecker,
	S.~L. Portalupi, T.~Grange, C.~Anton, J.~Demory, et~al., Near-optimal
	single-photon sources in the solid state, Nature Photonics 10~(5) (2016)
	340--345.
	
	\bibitem{gudmundsson2015coupled}
	V.~Gudmundsson, A.~Sitek, P.-y. Lin, N.~R. Abdullah, C.-S. Tang, A.~Manolescu,
	Coupled collective and Rabi oscillations triggered by electron transport
	through a photon cavity, ACS Photonics 2~(7) (2015) 930--934.
	
	\bibitem{childress2004mesoscopic}
	L.~Childress, A.~S{\o}rensen, M.~D. Lukin, Mesoscopic cavity quantum
	electrodynamics with quantum dots, Physical Review A 69~(4) (2004) 042302.
	
	\bibitem{shibata2012photon}
	K.~Shibata, A.~Umeno, K.~Cha, K.~Hirakawa, Photon-assisted tunneling through
	self-assembled InAs quantum dots in the terahertz frequency range, Physical
	Review Letters 109~(7) (2012) 077401.
	
	\bibitem{gustavsson2007frequency}
	S.~Gustavsson, M.~Studer, R.~Leturcq, T.~Ihn, K.~Ensslin, D.~Driscoll,
	A.~Gossard, Frequency-selective single-photon detection using a double
	quantum dot, Physical Review Letters 99~(20) (2007) 206804.
	
	\bibitem{Efficient342017}
	T.~H. Jonsson, A.~Manolescu, H.-S. Goan, N.~R. Abdullah, A.~Sitek, C.-S. Tang,
	V.~Gudmundsson, \href{http://dx.doi.org/10.1016/j.cpc.2017.06.018}{Efficient
		determination of the markovian time-evolution towards a steady-state of a
		complex open quantum system}, Computer Physics Communications 220 (2017)
	81–90.
	\newblock \href {https://doi.org/10.1016/j.cpc.2017.06.018}
	{\path{doi:10.1016/j.cpc.2017.06.018}}.
	\newline\urlprefix\url{http://dx.doi.org/10.1016/j.cpc.2017.06.018}
	
	\bibitem{ABDULKHALAQ2022414097}
	H.~A. Abdulkhalaq, N.~R. Abdullah, V.~Gudmundsson,
	\href{https://www.sciencedirect.com/science/article/pii/S0921452622004033}{Effects
		of coupling strength of the electron-photon and the photon-environment
		interactions on the electron transport through multiple-resonances of a
		double quantum dot system in a photon cavity}, Physica B: Condensed Matter
	(2022) 414097\href
	{https://doi.org/https://doi.org/10.1016/j.physb.2022.414097}
	{\path{doi:https://doi.org/10.1016/j.physb.2022.414097}}.
	\newline\urlprefix\url{https://www.sciencedirect.com/science/article/pii/S0921452622004033}
	
	\bibitem{Rabi-resonant-abdullah2019}
	N.~R. Abdullah, Rabi-resonant and intraband transitions in a multilevel quantum
	dot system controlled by the cavity-photon reservoir and the electron-photon
	coupling, Results in Physics 15 (2019) 102686.
	
	\bibitem{2019b}
	V.~Gudmundsson, N.~R. Abdullah, C.~Tang, A.~Manolescu, V.~Moldoveanu,
	\href{http://dx.doi.org/10.1002/andp.201900306}{Cavity‐photon‐induced
		high‐order transitions between ground states of quantum dots}, Annalen der
	Physik 531~(11) (2019) 1900306.
	\newblock \href {https://doi.org/10.1002/andp.201900306}
	{\path{doi:10.1002/andp.201900306}}.
	\newline\urlprefix\url{http://dx.doi.org/10.1002/andp.201900306}
	
	\bibitem{Stepwise442012}
	V.~Gudmundsson, O.~Jonasson, T.~Arnold, C.-S. Tang, H.-S. Goan, A.~Manolescu,
	\href{http://dx.doi.org/10.1002/prop.201200053}{Stepwise introduction of
		model complexity in a generalized master equation approach to time-dependent
		transport}, Fortschritte der Physik 61~(2-3) (2012) 305–316.
	\newblock \href {https://doi.org/10.1002/prop.201200053}
	{\path{doi:10.1002/prop.201200053}}.
	\newline\urlprefix\url{http://dx.doi.org/10.1002/prop.201200053}
	
	\bibitem{nakajima1958quantum}
	S.~Nakajima, On quantum theory of transport phenomena: Steady diffusion,
	Progress of Theoretical Physics 20~(6) (1958) 948--959.
	
	\bibitem{zwanzig1960ensemble}
	R.~Zwanzig, Ensemble method in the theory of irreversibility, The Journal of
	Chemical Physics 33~(5) (1960) 1338--1341.
	
	\bibitem{correlations45b2018}
	V.~Gudmundsson, N.~R. Abdullah, A.~Sitek, H.-S. Goan, C.-S. Tang, A.~Manolescu,
	\href{http://dx.doi.org/10.1016/j.physleta.2018.04.017}{Current correlations
		for the transport of interacting electrons through parallel quantum dots in a
		photon cavity}, Physics Letters A 382~(25) (2018) 1672–1678.
	\newblock \href {https://doi.org/10.1016/j.physleta.2018.04.017}
	{\path{doi:10.1016/j.physleta.2018.04.017}}.
	\newline\urlprefix\url{http://dx.doi.org/10.1016/j.physleta.2018.04.017}
	
	\bibitem{vidar_2021}
	V.~Gudmundsson, N.~R. Abdullah, C.-S. Tang, A.~Manolescu, V.~Moldoveanu,
	\href{http://dx.doi.org/10.1016/j.physe.2020.114544}{Self-induction and
		magnetic effects in electron transport through a photon cavity}, Physica E:
	Low-dimensional Systems and Nanostructures 127 (2021) 114544.
	\newblock \href {https://doi.org/10.1016/j.physe.2020.114544}
	{\path{doi:10.1016/j.physe.2020.114544}}.
	\newline\urlprefix\url{http://dx.doi.org/10.1016/j.physe.2020.114544}
	
	\bibitem{gudmundsson2017regimes}
	V.~Gudmundsson, T.~H. Jonsson, M.~L. Bernodusson, N.~R. Abdullah, A.~Sitek,
	H.-S. Goan, C.-S. Tang, A.~Manolescu, Regimes of radiative and nonradiative
	transitions in transport through an electronic system in a photon cavity
	reaching a steady state, Annalen der Physik 529~(1-2) (2017) 1600177.
	
	\bibitem{hohenester2010cavity}
	U.~Hohenester, Cavity quantum electrodynamics with semiconductor quantum dots:
	Role of phonon-assisted cavity feeding, Physical Review B 81~(15) (2010)
	155303.
	
\end{thebibliography}

\end{document}